\documentclass[intlimits,twoside,a4paper]{article}

\usepackage{amsmath,amssymb}
\usepackage{graphicx}
\usepackage{wrapfig}
\usepackage{color}

\usepackage[T2A]{fontenc}
\usepackage[cp1251]{inputenc}
\usepackage[eqsecnum]{cmpj2}

\issue{2013}{16}{2}{23004}

\doinumber{10.5488/CMP.16.23004}

\title[Mechanism of collisionless sound damping in dilute Bose gas with condensate]%
{Mechanism of collisionless sound damping  \\ in dilute Bose gas with condensate}%
\author[Yu. Slyusarenko, A. Kruchkov]{Yu. Slyusarenko\refaddr{label1,label2}\thanks{E-mail: slusarenko@kipt.kharkov.ua}\,,   A.  Kruchkov \refaddr{label2}\thanks{E-mail: aleks.kryuchkov@gmail.com}}
\addresses{
\addr{label1} Akhiezer Institute for Theoretical Physics, National Science Center Kharkiv Institute of Physics and Technology, 1~Akademichna St., 61108 Kharkiv, Ukraine
\addr{label2} Karazin National University, 4~Svobody Sq., 61077 Kharkiv, Ukraine
}

\date{Received November 24, 2012, in final form March 18, 2013}
\authorcopyright{Yu. Slyusarenko, A. Kruchkov, 2013}

\begin{document}

\sloppy

\maketitle

\begin{abstract}
We develop a microscopic theory of sound damping due to Landau mechanism in dilute gas with Bose condensate. It is based on the coupled evolution equations of the parameters describing the system. These equations have been derived in earlier works within a microscopic approach which employs the Peletminskii-Yatsenko reduced description method for quantum many-particle systems and Bogoliubov model for a weakly nonideal Bose gas with a separated condensate. The dispersion equations for sound oscillations were obtained by linearization of the mentioned evolution equations in the collisionless approximation. They were analyzed both analytically and numerically. The expressions for sound speed and decrement rate  were obtained in high and low temperature limiting cases. We have shown that at low temperature the dependence of the obtained quantities on temperature significantly differs from those obtained by other authors in the semi-phenomenological approaches. Possible effects connected with non-analytic temperature dependence of dispersion characteristics of the system were also indicated.

\pacs 05.30.-d, 05.30.Jp, 67.85.Hj, 67.85.Jk, 03.75.Hh, 03.75.Kk
\end{abstract}

\section{Introduction}

The study of mechanisms of sound damping in a Bose-Einstein condensate (BEC) has a long history. Calculation of the sound damping rate in systems with BEC is a rather complicated theoretical problem. First expressions for damping rate in such systems have been apparently obtained in \cite{1, 2} for the spatially homogeneous case.

The direct experimental observation of BEC \cite{3,4,5}  has stimulated a great number of works devoted to various aspects of this phenomenon (see, for example, \cite{6, 7} and references therein). A number of papers, both theoretical and experimental, deal with the problem of  propagation and damping of excitations in Bose gases with the presence of condensate \cite{8,9,10,11,12a,12b,13,14,15,16}.

It is currently assumed that the Landau damping is the most probable mechanism of sound relaxation in the so-called trapped Bose condensates. This mechanism consists in collisionless absorption of oscillation energy by quanta of elementary excitations \cite{12a,12b,13}. In this regard, recall that the existence of specific collective excitations in a gas with BEC has been known since the pioneering work of Bogoliubov \cite{17}. In this paper, a special perturbation theory was proposed for a weakly non-ideal and spatially homogeneous Bose gas with condensate in which the repulsive interaction acts between atoms. This theory predicts the elementary excitation spectrum for such system at zero temperature. At small wave vectors, it coincides with the spectrum of sound oscillations in a condensed Bose gas.

The first work, which proposed a method to calculate the sound damping rate in trapped BEC due to Landau mechanism, is apparently the paper \cite{12a} (see also \cite{13}). The approach developed by the authors uses the perturbation theory and is based on the calculation of the difference in probabilities between emission and absorption of quanta of oscillations by elementary excitations in a system. It should be noted that trapped condensates represent a spatially inhomogeneous system. This fact essentially complicates analytical calculations. The method of \cite{12a,12b} was shown to be suitable for numerical calculations of sound damping in a trapped condensate (see the same paper). For a spatially homogeneous BEC, this method has provided analytical formulae for damping according to Landau mechanism \cite{12a,12b}. In this case, the authors  reproduced the results obtained in \cite{1,2}.

It should be stressed that the formulae of \cite{12a,12b} (and, hence, the results of \cite{1,2}) can be obtained by another method involving the kinetic equation for distribution function of elementary excitations (see, e.g., \cite{6}). In other words, the semi-phenomenological approach to the calculation of the damping rate, shown in \cite{6}, is equivalent to the method of \cite{12a,12b}. It was inter alia indicated in \cite{12a,12b}. However, it is clear that in the most general case, the mere use of kinetic equation for excitations is insufficient. The system should be described by the coupled evolution equations which take into account the mutual effect of condensate density, phase (or superfluid velocity) and distribution function of elementary excitations. A consistent derivation of a system of coupled equations can be achieved using a microscopic approach, proceeding from the first principles. This problem was solved in \cite{18,19,10,21} within a microscopic approach based on the reduced description of quantum many-particle systems \cite{22,23} and Bogoliubov model for a weakly non-ideal Bose gas with condensate \cite{17}. The synthesis of the approaches elaborated in \cite{17} and \cite{23} made it possible to obtain in \cite{18,19} the kinetic equation for distribution function of elementary excitations coupled with the evolution equations for condensate density and superfluid momentum. Note that the validity of such a system of equations is confirmed by its controlled derivation within the framework of special perturbation theory with weak interparticle interaction. Furthermore, the following fact speaks in favour of the mentioned coupled equations: they have been employed in \cite{20,21} to derive hydrodynamic equations of a superfluid in which the smallness of the difference between superfluid and normal velocities is not taken into account. As this difference tends to zero, the obtained equations are reduced to the well-known Khalatnikov hydrodynamic equations (see e.g. \cite{24}).  Also note that evolution equations from [18, 19] were in fact reproduced in \cite{Kirkpatrick} in another microscopic approach. This circumstance was mentioned in \cite{Kirkpatrick} with an appropriate reference link.

Notwithstanding the above considerations, the equations of \cite{19,Kirkpatrick} have not yet been used to study the propagation and damping of sound in a dilute gas with BEC. However, the same considerations allow us to hope that the correct solution of evolution equations found in \cite{19} should lead us to the correct expression for sound damping rate in a gas with BEC. The present paper is devoted to the study of a collisionless mechanism of sound damping in dilute gases with BEC on the basis of general dynamic equations of such systems obtained in \cite{19} from the first principles. As will be seen later, the results obtained in the present work significantly differ in some cases from those of \cite{1,2} and, consequently, of \cite{12a,12b} (see also \cite{Giorgini,Jackson,Escobedo}). For example, in the present study it is shown that in the collisionless approximation the sound damping rate at low temperature is quadratic on temperature, $\gamma \sim T^{2} $, whereas the results of \cite{1,12a,12b} give $\gamma \sim T^{4} $ dependence.

\section{Kinetics of spatially inhomogeneous Bose gas with the presence of condensate}

Constructing spatially inhomogeneous Bose gas kinetics, authors of \cite{19} started from the Liouville equation for the statistical operator $\rho (t)$
\begin{equation} \label{GrindEQ__1_1_}
i\frac{\partial \rho \left(t\right)}{\partial t} =[\hat{H},\rho \left(t\right)],
\end{equation}
where $\hat{H}=\hat{H}_{0} +\hat{V}$ is Hamiltonian of the system, consisting of the ideal gas Hamiltonian $\hat{H}_{0} $
\begin{equation} \label{GrindEQ__1_2_}
\hat{H}_{0} =\frac{1}{2m} \int \rd^{3} r\frac{\partial \hat{\psi }^{+} \left(\mathbf{r}\right)}{\partial r_{k} } \frac{\partial \hat{\psi }\left(\mathbf{r}\right)}{\partial r_{k} }
\end{equation}
and binary interaction Hamiltonian $\hat{V}$
\begin{equation}\begin{split}\begin{gathered} \label{GrindEQ__1_3_}
\hat{V}=\frac{1}{2} \int d^{3} r \int d^{3} R \hat{\psi }^{+} \left(\mathbf{r}+\mathbf{R}\right)\hat{\psi }^{+} \left(\mathbf{r}\right)V\left(\left|\mathbf{R}\right|\right) \hat{\psi }\left(\mathbf{r}\right)\hat{\psi }\left(\mathbf{r}+\mathbf{R}\right).
\end{gathered}\end{split}\end{equation}

Equation \eqref{GrindEQ__1_1_} is written in the units in which Planck's constant $\hbar $ is equal to unity. In the formulae \eqref{GrindEQ__1_2_}, \eqref{GrindEQ__1_3_} \textit{m} is boson mass, $V\left(\left|\mathbf{R}\right|\right)$ is  binary interaction potential, which depends only on the distance between particles, and $\hat{\psi }^{+} \left(\mathbf{r}\right),\, \hat{\psi }\left(\mathbf{r}\right)$ are field operators. The quasiparticle distribution function $f_{\mathbf{p}} \left(\mathbf{r},t\right)$, the order parameter $\psi \left(\mathbf{r},t\right)=|\mathrm{Sp} \rho (t)\hat{\psi }\left(\mathbf{r}\right)|$ and the superfluid velocity $v_{k} \left(\mathbf{r},t\right)=m^{-1} \frac{\partial }{\partial r_{k} } \Im\ln \mathrm{Sp}\rho (t)\hat{\psi }\left(\mathbf{r}\right)$ were selected as parameters to describe  weakly non-ideal Bose gas with condensate in kinetic stage of system evolution in [19].

Using the method of reduced description \cite{23} combined with the special perturbation theory \cite{17} made it possible to obtain in \cite{19} the following system of equations for the parameters $f_{\mathbf{p}} \left(\mathbf{r},t\right)$, $\psi \left(\mathbf{r},t\right)$ and $v_{k} \left(\mathbf{r},t\right)$. Since the general form of equations from \cite{19} is not required in the present paper,  we introduce it here in collisionless approximation:
\begin{eqnarray}
 \label{GrindEQ__1_9_}
&&\frac{\partial f_{\mathbf{p}} }{\partial t}
+\frac{\partial f_{\mathbf{p}} }{\partial r_{k} } \frac{\partial }{\partial p_{k} } \varepsilon _{\mathbf{p}} \left(n,\mathbf{v}\right)
-\frac{\partial f_{\mathbf{p}} }{\partial p_{k} } \frac{\partial }{\partial r_{k} } \varepsilon _{\mathbf{p}} \left(n,\mathbf{v}\right)+o\left( \lambda \right)=0,\nonumber
\\
&&\frac{\partial n}{\partial t} +\frac{\partial }{\partial r_{k} } \left(v_{k} n\right)=0,
\qquad
 \frac{\partial v_{k} }{\partial t} +\frac{\partial }{\partial r_{k} } \left\{\frac{v^{2} }{2} +h\left(f,n\right)\right\}=0,
\end{eqnarray}
where instead of description parameter  $\psi(\mathbf{r},t)$ we have introduced a new variable $n(\mathbf{r},t)$, that is condensate density \cite{21}
\begin{equation}
 \label{GrindEQ__1_10_}
\psi ^{2} (\mathbf{r},t)\equiv n(\mathbf{r},t),
\end{equation}
and the quantities $\varepsilon _{\mathbf{p}}\left(n,\mathbf{v}\right)$ and $h\left(f,n \right)$ are given by expressions:
\begin{eqnarray}
\label{GrindEQ__1_11_}
&&\varepsilon _{\mathbf{p}} \left(n,\mathbf{v}\right)\equiv \omega _{\mathbf{p}} \left(n\right)+\mathbf{p}\cdot\mathbf{v},
 \qquad
h\left(f,n\right)=
\frac{1}{m} \left\langle \nu _{\mathbf{p}} \frac{\varepsilon _{\mathbf{p}} }{\omega _{\mathbf{p}} } \right\rangle
+\frac{\nu _{0} }{2m} \left\langle \frac{\alpha _{\mathbf{p}} }{\omega _{\mathbf{p}} } \right\rangle
 +h_{0} \left(n\right)+o\left( \lambda^{4} \right),\nonumber
\\[1ex]
&&h_{0} \left(n\right)=
\frac{1}{V} \sum _{\mathbf{p}\ne 0}\frac{1}{2m}  \left(\nu _{\mathbf{p}} \frac{\varepsilon _{\mathbf{p}} -\omega _{\mathbf{p}} }{\omega _{\mathbf{p}} }
+\nu _{0} \frac{\alpha _{\mathbf{p}} -\omega _{\mathbf{p}} }{\omega _{\mathbf{p}} } \right)
+\frac{\nu _{0} n}{m} \,,
\end{eqnarray}
\noindent where
\begin{equation}\begin{split}\begin{gathered}
 \label{GrindEQ__1_12_}
\omega _{\mathbf{p}} \left(n\right)\equiv \left\{\varepsilon _{p} \left[\varepsilon _{p} +2\beta _{\mathbf{p}} \left(n\right)\right]\right\}^{1/2} \,,
\qquad
  \alpha _{\mathbf{p}} \left(n\right)\equiv \varepsilon _{p} +\beta _{\mathbf{p}} \left(n\right)\,,
\qquad
\varepsilon _{p} \equiv \frac{p^{2} }{2m} \,,
\quad
 \beta _{\mathbf{p}} \left(n\right)\equiv \nu _{\mathbf{p}} n  \,,
\end{gathered}\end{split}\end{equation}
\noindent and the following notations were also introduced
\begin{equation}\begin{split}\begin{gathered}
 \label{GrindEQ__1_6_}
\left\langle A_{\mathbf{p}} \right\rangle \equiv \frac{1}{\left(2 \pi \right)^{3}} \int \rd^{3} pf_{\mathbf{p}}  A_{\mathbf{p}} \,,
 \ \ \
 \nu _{0} \equiv \nu _{\mathbf{p}=0}\, ,
\ \ \
 \nu _{\mathbf{p}} \equiv \int \rd^{3} R \, V\left(\left|\mathbf{R}\right|\right)\exp \left(-\ri\mathbf{p}\cdot\mathbf{R}\right),
\end{gathered}\end{split}\end{equation}
\noindent where $A_{\mathbf{p}}$ is an arbitrary function of $\mathbf{p}$.

Let us recall that the basis of Bogoliubov equilibrium state theory of Bose gas in the presence of interaction is the assumption that $\psi \sim \lambda ^{-1} $, see \cite{17}, where quantity $\lambda$ characterizes the smallness of the interaction between the particles, $V\left(\left|\mathbf{R}\right|\right)\sim \lambda ^{2} $. Furthermore, it is believed that the order of magnitude of $\psi \left(r,t\right)$ does not change after differentiation $\psi \left(\mathbf{r},t\right)$ with respect to $\mathbf{r}$, $\partial \psi \left(\mathbf{r},t\right)/\partial r_{k}\sim \lambda ^{-1} $. In the formulae \eqref{GrindEQ__1_9_}--\eqref{GrindEQ__1_11_}, the notation  $o\left(\lambda ^{n} \right)$  means the quantity in order of magnitude of $\lambda ^{n}$.

We emphasize once again that in our paper we study the mechanism of collisionless sound damping in a gas with a BEC (Landau mechanism). For this reason, in  equations \eqref{GrindEQ__1_9_}--\eqref{GrindEQ__1_11_} we have omitted the terms associated with the presence of interparticle collision term. The explicit form of the collision term for quasiparticles can be found in \cite{18, 19}. Here, we note only the fact that the quasiparticle collision term $L_{\mathbf{p}} \left(f,\psi \right)$ vanishes by substituting the stationary Bose distribution function $f_{\mathbf{p}}^{0} $
\begin{equation}\begin{split} \label{GrindEQ__1_8_}
L_{\mathbf{p}} \left(f^{0} ,\psi \right)=0\,,
\qquad
  f_{\mathbf{p}}^{0} =\left[\exp\left( \frac{\omega _{\mathbf{p}} -\mathbf{p}\cdot\mathbf{v}}{T}\right) -1\right]^{-1},
\end{split}\end{equation}
\noindent  with the chemical potential of quasiparticles being equal to zero; here $T$ is temperature in energy units. The chemical potential having vanished reflects the fact that the number of quasiparticles is not conserved during collisions \cite{18,21}.

\section{Sound dispersion equations in diluted gas with BEC}

To investigate the  propagation of sound in gas with BEC, we linearize coupled equations \eqref{GrindEQ__1_9_}, \eqref{GrindEQ__1_11_}, \eqref{GrindEQ__1_12_} with respect to spatially homogeneous equilibrium state according to the following formulae
\begin{eqnarray}
\label{GrindEQ__2_1_}
&n(\mathbf{r},t)=n_{0} +\tilde{n}(\mathbf{r},t),
\qquad  \mathbf{v}(\mathbf{r},t)=\tilde{\mathbf{v}}(\mathbf{r},t),
\qquad
f_{\mathbf{p}} (\mathbf{r},t)=f_{\mathbf{p}}^{0} +\tilde{f}_{\mathbf{p}} (\mathbf{r},t),\nonumber
&\\
&n_{0} \gg |\tilde{n}(\mathbf{r},t)|,
\qquad
f_{\mathbf{p}}^{0} \gg |\tilde{f}_{\mathbf{p}} (\mathbf{r},t)|,&
\end{eqnarray}
\noindent  where $n_{0} =n_{0} \left(T\right)$ is the equilibrium value of the condensate density in the system at the temperature $T$. The equilibrium value of the velocity $\mathbf{v}_{0} $ is considered to be equal to zero in the second formula of \eqref{GrindEQ__2_1_}. Thus, the velocity $\left|\tilde{\mathbf{v}}(\mathbf{r},t)\right|$ is supposed to be of the order of magnitude of $\tilde{n}(\mathbf{r},t)$ and $\tilde{f}_{\mathbf{p}} (\mathbf{r},t)$. Then, the equilibrium distribution function of quasiparticles pursuant to \eqref{GrindEQ__1_8_} is given by the expression
\begin{equation} \label{GrindEQ__2_2_}
f_{\mathbf{p}}^{0} =\left[\exp\left( \frac{\omega _{\mathbf{p}} }{T}\right) -1\right]^{-1} .
\end{equation}

In this formula, like in all subsequent expressions, we omit the index ``0'' in the designation of the equilibrium value of $\omega _{\mathbf{p}}^{0} \equiv \omega _{\mathbf{p}} \left(n_{0} \right)$ to avoid encumbering the computations. Note that the quantity $h\left(f^{0} ,n_{0} \right)$defined by \eqref{GrindEQ__1_11_} represents the chemical potential of atomic Bose gas $\mu =h\left(f^{0} ,n_{0} \right)$ (see in this regard \cite{18,19}).

Deviations of the corresponding quantities from their equilibrium values were denoted by $\tilde{n}(\mathbf{r},t)$,
$\tilde{\mathbf{v}}(\mathbf{r},t)$
and $\tilde{f}_{\mathbf{p}} (\mathbf{r},t)$.

The equations of motion for these variables can be represented as follows:
\begin{eqnarray}
 \label{GrindEQ__2_3_}
&&\frac{\partial }{\partial t} \tilde{n}(\mathbf{r},t)
+n_{0} \frac{\partial }{\partial r_{k} } \tilde{v}_{k} (\mathbf{r},t)=0\,,\nonumber
 \\
&& \frac{\partial }{\partial t} \tilde{v}_{k} (\mathbf{r},t)
+\frac{\nu _{0} }{m} \frac{\partial }{\partial r_{k} } \tilde{n}(\mathbf{r},t)
+ \int \rd^{3} p K_{\mathbf{p}}  \frac{\partial }{\partial r_{k} } \tilde{f}_{\mathbf{p}} (\mathbf{r},t)=0\,,\nonumber
\\
&&\frac{\partial }{\partial t} \tilde{f}_{\mathbf{p}} (\mathbf{r},t)
+\frac{\partial \omega _{\mathbf{p}} }{\partial p_{k} } \frac{\partial }{\partial r_{k} } \tilde{f}_{\mathbf{p}} (\mathbf{r},t)
-\frac{\partial f_{\mathbf{p}}^{0} }{\partial p_{k} }
\frac{\partial }{\partial r_{k} } \left[\mathbf{p}\cdot\tilde{\mathbf{v}}(\mathbf{r},t)
+\frac{\nu _{\mathbf{p}} \varepsilon _{p} }{\omega _{\mathbf{p}} } \tilde{n}(\mathbf{r},t)\right]=0\,,
\end{eqnarray}
\noindent  where we have introduced the following notation
\begin{equation} \label{GrindEQ__2_4_}
K_{\mathbf{p}} =
\frac{1}{2m\left(2\pi \right)^{3} } \frac{\varepsilon _{p} \left(\nu _{0} +2\nu _{\mathbf{p}} \right)+\nu _{\mathbf{p}} \nu _{0} n_{0} }{\omega _{\mathbf{p}} } \,.
\end{equation}

Recall that in the present paper, the Planck constant $\hbar $ is considered to be equal to unity. Note also that while deriving equations \eqref{GrindEQ__2_3_}, we have discarded the terms that are higher than the first order of magnitude $\lambda $. The circumstance of \eqref{GrindEQ__1_10_} stating that $n_{0} \sim \lambda ^{-2} $ or i.e. $\nu _{\mathbf{p}} n_{0} \sim \lambda ^{0} $ was taken into account.

Proceeding further to the Fourier transform of the quantities
$\tilde{n}(\mathbf{r},t)$,
$\tilde{\mathbf{v}}(\mathbf{r},t)$,
$\tilde{f}_{\mathbf{p}} (\mathbf{r},t)$
in equations \eqref{GrindEQ__2_3_} as provided by formula
\begin{equation} \label{GrindEQ__2_5_}
\tilde{\zeta }(\mathbf{r},t)=\int _{-\infty }^{\infty }\rd\omega \int \rd^{3} q\,\re^{\ri\mathbf{q}\mathbf{r}-\ri\omega t}   \zeta (\mathbf{q},\omega ),
\end{equation}
\noindent  where $\tilde{\zeta }(\mathbf{r},t)$ should be understood as either of the considered quantities, we obtain
\begin{align}
 \label{GrindEQ__2_6_}
\omega n(\mathbf{q},\omega )&=n_{0} \mathbf{q}\cdot\mathbf{v}\left(\mathbf{q},\omega \right),
\nonumber \\
\omega v_{k} (\mathbf{q},\omega )&=\frac{\nu _{0} }{m} q_{k} n\left(\mathbf{q},\omega \right)+q_{k} A\left(\mathbf{q},\omega \right),\nonumber
 \\
\left(\omega -q_{k} \frac{\partial \omega _{\mathbf{p}} }{\partial p_{k} } \right)f_{\mathbf{p}} \left(\mathbf{q},\omega \right)&=
-q_{k} \frac{\partial f_{\mathbf{p}}^{0} }{\partial p_{k} } \left[\mathbf{p}\cdot\mathbf{v}\left(\mathbf{q},\omega \right)+\frac{\nu _{\mathbf{p}} \varepsilon _{p} }{\omega _{\mathbf{p}} } n\left(\mathbf{q},\omega \right)\right],
\end{align}
\noindent  where we denote [see \eqref{GrindEQ__2_3_}, \eqref{GrindEQ__2_4_}]
\begin{equation} \label{GrindEQ__2_7_}
A\left(\mathbf{q},\omega \right)\equiv \int \rd^{3} p f_{\mathbf{p}} \left(\mathbf{q},\omega \right)K_{\mathbf{p}}
= \frac{1}{2m\left(2\pi \right)^{3} } \int \rd^{3} p f_{\mathbf{p}} \left(q,\omega \right)\frac{\varepsilon _{p} \left(\nu _{0} +2\nu _{\mathbf{p}} \right)+\nu _{\mathbf{p}} \nu _{0} n_{0} }{\omega _{\mathbf{p}} }\, .
\end{equation}

In the formulae \eqref{GrindEQ__2_6_}, \eqref{GrindEQ__2_7_} in Fourier transforms of the quantities
$\tilde{n}(\mathbf{r},t)$,
$\tilde{\mathbf{v}}(\mathbf{r},t)$,
$\tilde{f}_{\mathbf{p}} (\mathbf{r},t)$
we omit the 'tilde' sign.
Further expressing the values $n(\mathbf{q},\omega )$ and $\mathbf{v}\left(\mathbf{q},\omega \right)$ from the first two equations of \eqref{GrindEQ__2_6_} in terms of $A\left(\mathbf{q},\omega \right)$,
\begin{equation}\begin{split}  \label{GrindEQ__2_8_}
n(\mathbf{q},\omega )=\frac{n_{0} q^{2} }{\omega ^{2} -u_{0}^{2} q^{2} } A\left(\mathbf{q},\omega \right),
\qquad
v_{k} (\mathbf{q},\omega )=\frac{\omega q_{k} }{\omega ^{2} -u_{0}^{2} q^{2} } A\left(\mathbf{q},\omega \right),
\end{split}\end{equation}
\noindent  the third equation of \eqref{GrindEQ__2_6_} can be written in the form
\begin{equation}
\label{GrindEQ__2_9_}
\left(\omega -q_{k} \frac{\partial \omega _{\mathbf{p}} }{\partial p_{k} } \right)f_{\mathbf{p}} \left(\mathbf{q},\omega \right)=
-q_{k} \frac{\partial \omega _{\mathbf{p}} }{\partial p_{k} } \frac{\partial f_{\mathbf{p}}^{0} }{\partial \omega _{\mathbf{p}} }
\frac{\mathbf{p}\cdot\mathbf{q}
\omega \omega _{\mathbf{p}} +\nu _{\mathbf{p}} \varepsilon _{p} n_{0} q^{2} }{\omega _{\mathbf{p}} \left(\omega ^{2} -u_{0}^{2} q^{2} \right)} A\left(\mathbf{q},\omega \right),
\end{equation}
\noindent  where $u_{0}^{} $ is referred to as speed of zero sound in Bose gas
\begin{equation} \label{GrindEQ__2_10_}
u_{0}^{2} =\frac{\nu _{0} n_{0} }{m} \,.
\end{equation}

As is readily seen, equation \eqref{GrindEQ__2_9_} subject to \eqref{GrindEQ__2_7_} is an integral equation for distribution function Fourier transform $f_{\mathbf{p}}\left(\mathbf{q},\omega \right)$. The solution of this equation can be represented as in \cite{25}:
\begin{equation}\begin{split}\begin{gathered}
 \label{GrindEQ__2_11_}
f_{\mathbf{p}} \left(\mathbf{q},\omega \right)
=B_{\mathbf{p}} \left(\mathbf{q}\right)\delta \left(\omega -\mathbf{q}\cdot\mathbf{u}_{p} \right)
 -\mathbf{q}\cdot\mathbf{{u}}_{p}
 \frac{\partial f_{\mathbf{p}}^{0} }{\partial \omega _{\mathbf{p}} } \frac{\mathbf{p}\cdot\mathbf{q}\omega \omega _{\mathbf{p}}
+\nu _{\mathbf{p}} \varepsilon _{p} n_{0} q^{2} }{\omega _{\mathbf{p}}
\left(\omega ^{2} -u_{0}^{2} q^{2} \right)\left(\omega -\mathbf{q}\cdot\mathbf{u}_{p} +\ri0\right)} A\left(\mathbf{q},\omega \right),
\end{gathered}\end{split}\end{equation}
\noindent  where we have introduced the following notation:
\begin{equation} \label{GrindEQ__2_12_}
\mathbf{u}_{p} \equiv \frac{\partial \omega _{\mathbf{p}} }{\partial \mathbf{p}} ,
\end{equation}
\noindent and $B_{\mathbf{p}} \left(\mathbf{q}\right)$ is an arbitrary function, which is required to impose the following restriction: the distribution function $\tilde{f}_{\mathbf{p}} (\mathbf{r},t)$,
calculated in accordance with \eqref{GrindEQ__2_1_}, \eqref{GrindEQ__2_5_} and \eqref{GrindEQ__2_11_} should be small compared with the equilibrium distribution function $f_{\mathbf{p}}^{0} $.

It implies also from \eqref{GrindEQ__2_5_} that $B_{\mathbf{p}} \left(\mathbf{q}\right)$ must satisfy the following relation:
\begin{equation} \label{GrindEQ__2_13_}
B_{\mathbf{p}}^{*} \left(\mathbf{q}\right)=B_{\mathbf{p}} \left(-\mathbf{q}\right).
\end{equation}

We denote the whole valid set of such functions by $B_{\sigma \mathbf{p}} \left(\mathbf{q}\right)$, where $\sigma $ is a continuous or discrete symbolic parameter such that the functions $B_{\mathbf{p}} \left(\mathbf{q}\right)\equiv B_{\sigma \mathbf{p}} \left(\mathbf{q}\right)$ may depend on $\sigma $. The reason for introducing this index may consist, for example, in the following: the set of functions $B_{\sigma \mathbf{p}} \left(\mathbf{q}\right)$ should be sufficient to build an arbitrary value for the distribution function $\tilde{f}_{\mathbf{p}} (\mathbf{r},t)$ at the initial time $t=0$ (in this context see also \cite{26}).

Formula \eqref{GrindEQ__2_11_} permits to find the value $A\left(\mathbf{q},\omega \right)$ in terms of $B_{\sigma \mathbf{p}} \left(\mathbf{q}\right)$ functions:
\begin{equation} \label{GrindEQ__2_14_}
A_{\sigma } \left(\mathbf{q},\omega \right)
=B_{\sigma } \left(\mathbf{q},\omega \right)\varepsilon ^{-1} \left(\mathbf{q},\omega \right),
\end{equation}
\noindent where
\begin{equation}
 \label{GrindEQ__2_15_}
B_{\sigma } \left(\mathbf{q},\omega \right)
=\frac{1}{2m\left(2\pi \right)^{3} }
\int \rd^{3} p \frac{\nu _{0} n_{0} \nu _{\mathbf{p}} +\varepsilon _{p} \left(\nu _{0} +2\nu _{\mathbf{p}} \right)}{\omega _{\mathbf{p}} }
B_{\sigma \mathbf{p}} \left(\mathbf{q}\right) \delta \left(\omega -\mathbf{q}\cdot\mathbf{u}_{p} \right)
\end{equation}
\noindent  and
\begin{equation}
\label{GrindEQ__2_16_}
\varepsilon \left(\mathbf{q},\omega \right)\equiv 1
+\frac{1}{2m\left(2\pi \right)^{3} }
\int \rd^{3} p \frac{\partial f_{\mathbf{p}}^{0} }{\partial \omega _{\mathbf{p}} }
 \left(\mathbf{q}\cdot\mathbf{p}\omega \omega _{\mathbf{p}}
+\nu _{\mathbf{p}} \varepsilon _{p} n_{0} q^{2} \right)
\frac{\mathbf{q}\cdot\mathbf{u}_{p}
\left[\nu _{0} n_{0} \nu _{\mathbf{p}} +\varepsilon _{p} \left(\nu _{0} +2\nu _{\mathbf{p}} \right)\right]}{\omega _{\mathbf{p}}^{2} \left(\omega ^{2} -u_{0}^{2} q^{2} \right)
\left(\omega -\mathbf{q}\cdot\mathbf{u}_{p} +\ri0\right)} \,.
\end{equation}

The expression \eqref{GrindEQ__2_11_} for the given values $f_{\mathbf{p}} \left(\mathbf{q},\omega \right)$ subject to \eqref{GrindEQ__2_14_}--\eqref{GrindEQ__2_16_} can now be represented in the form:
\begin{equation}
\label{GrindEQ__2_17_}
f_{\sigma \mathbf{p}} \left(\mathbf{q},\omega \right)
=B_{\sigma \mathbf{p}} \left(\mathbf{q}\right)\delta \left(\omega -\mathbf{q}\cdot\mathbf{u}_{p} \right)+
\varepsilon ^{-1} \left(\mathbf{q},\omega \right)\mathbf{q}\cdot\mathbf{u}_{p} \frac{\partial f_{\mathbf{p}}^{0} }{\partial \omega _{\mathbf{p}} }
\frac{\mathbf{p}\cdot\mathbf{q}\omega \omega _{\mathbf{p}}
+\nu _{\mathbf{p}} \varepsilon _{p} n_{0} q^{2} }{\omega _{\mathbf{p}} \left(\omega ^{2} -u_{0}^{2} q^{2} \right)\left(\omega -\mathbf{q}\cdot\mathbf{u}_{p} +\ri 0\right)}
B_{\sigma } \left(\mathbf{q},\omega \right).
\end{equation}

We note that in case of charged particles gas, the quantity $\varepsilon \left(\mathbf{q},\omega \right)$ [see \eqref{GrindEQ__2_16_}]
\begin{equation} \label{GrindEQ__2_18_}
\varepsilon \left(\mathbf{q},\omega \right)=\varepsilon _{1} \left(\mathbf{q},\omega \right)+\ri\varepsilon _{2} \left(\mathbf{q},\omega \right)
\end{equation}
\noindent  represents complex dielectric permittivity of the system (see e.g., \cite{23}). It is known that the presence of an imaginary term in dielectric permittivity indicates the energy dissipation of electromagnetic waves with dispersion relation that should be obtained from the equation
\begin{equation} \label{GrindEQ__2_19_}
\varepsilon \big(\mathbf{q},\omega _{0} \left(\mathbf{q}\right)-\ri\gamma _{\mathbf{q}} \big)=0.
\end{equation}

Moreover, the wave decrement $\gamma _{\mathbf{q}} $ is determined by an imaginary part
$\varepsilon _{2} \left(\mathbf{q},\omega \right)$ of the value $\varepsilon \left(\mathbf{q},\omega \right)$.
For this reason, weakly damped oscillations in the system
\begin{equation} \label{GrindEQ__2_20_}
\left|\omega _{0} \left(\mathbf{q}\right)\right|\gg \gamma _{\mathbf{q}}
\end{equation}
\noindent  can exist if only
\begin{equation} \label{GrindEQ__2_21_}
\left|\varepsilon _{1} \left(\mathbf{q},\omega \right)\right|\gg \left|\varepsilon _{2} \left(\mathbf{q},\omega \right)\right|,
\end{equation}
\noindent besides, as the consequence of \eqref{GrindEQ__2_18_}--\eqref{GrindEQ__2_21_} (see in this regard \cite{Kirkpatrick,Giorgini}), the frequency $\omega _{0} \left(q\right)$ can be found from the equation
\begin{equation} \label{GrindEQ__2_22_}
\varepsilon _{1} \big(\mathbf{q},\omega _{0} \left(\mathbf{q}\right)\big)=0
\end{equation}
\noindent  and the damping rate $\gamma _{\mathbf{q}} $ is given by expression
\begin{equation} \label{GrindEQ__2_23_}
\gamma _{\mathbf{q}} =\left[\frac{\partial \varepsilon _{1} \left(\mathbf{q},\omega \right)}{\partial \omega } \right]_{\omega =\omega _{0} }^{-1} \varepsilon _{2} \big(\mathbf{q},\omega _{0} \left(q\right)\big).
\end{equation}

Despite the fact that in the present paper we investigate a neutral system, the existence of longitudinal oscillations is also associated with the existence of zeros of the function  $\varepsilon \left(\mathbf{q},\omega \right)$. In this case, there is a complete analogy with the mentioned case of longitudinal oscillations in systems of charged particles. That is, the structure of solution \eqref{GrindEQ__2_17_} is such that weakly damped waves may also be excited in the  system investigated, in accordance with \eqref{GrindEQ__2_8_} and \eqref{GrindEQ__2_5_}, and the dispersion law is determined by \eqref{GrindEQ__2_19_}--\eqref{GrindEQ__2_21_}. As we shall show in the following section, such waves would represent sound waves in weakly nonideal Bose gas with condensate. The value $\gamma _{\mathbf{q}} $ obtained according to \eqref{GrindEQ__2_17_}--\eqref{GrindEQ__2_20_} will determine the damping of sound in the system investigated.

\section{Sound damping rate in dilute Bose gas with condensate}

To solve this problem it is necessary to determine real $\varepsilon _{1} \left(\mathbf{q},\omega \right)$ and imaginary $\varepsilon _{2} \left(\mathbf{q},\omega \right)$
parts of the quantity $\varepsilon \left(\mathbf{q},\omega \right)$.
 For this purpose in the expression \eqref{GrindEQ__2_16_} we use formula
\begin{equation} \label{GrindEQ__3_1_}
\frac{1}{x+\ri0} =P\frac{1}{x} -\ri\pi \delta \left(x\right),
\end{equation}
\noindent  where the symbol $P$ means that the further integration is taken in the sense of Cauchy principal value. After some hackneyed transformations one can obtain the following expressions for
$\varepsilon _{1} \left(\mathbf{q},\omega \right)$ and $\varepsilon _{2} \left(\mathbf{q},\omega \right)$:
\begin{eqnarray}
 \label{GrindEQ__3_2_}
&&\varepsilon _{1} \left(\mathbf{q},\omega \right)=1+\frac{1}{2m\left(2\pi \right)^{3} } P\int \rd^{3} p \frac{\partial f_{\mathbf{p}}^{0} }{\partial \omega _{\mathbf{p}} }
\frac{\mathbf{q}\cdot\mathbf{u}_{p}
 \left(\mathbf{q}\cdot\mathbf{p}\omega \omega _{\mathbf{p}} +\nu _{\mathbf{p}} \varepsilon _{p} n_{0} q^{2} \right)
\left[\nu _{0} n_{0} \nu _{\mathbf{p}}
+\varepsilon _{p} \left(\nu _{0} +2\nu _{\mathbf{p}} \right)\right]}{\omega _{\mathbf{p}}^{2} \left(\omega ^{2} -u_{0}^{2} q^{2} \right)\left(\omega -\mathbf{q}\cdot\mathbf{u}_{p} \right)}\, ,\nonumber\\
&&\varepsilon _{2} \left(\mathbf{q},\omega \right)=-\frac{\pi \omega }{2m\left(2\pi \right)^{3} } \int \rd^{3} p \frac{\partial f_{\mathbf{p}}^{0} }{\partial \omega _{\mathbf{p}} }
\left(\mathbf{q}\cdot\mathbf{p}\omega \omega _{\mathbf{p}} +\nu _{\mathbf{p}} \varepsilon _{p} n_{0} q^{2} \right)
\frac{\left[\nu _{0} n_{0} \nu _{\mathbf{p}}
 +\varepsilon _{p} \left(\nu _{0} +2\nu _{\mathbf{p}} \right)\right]}{\omega _{\mathbf{p}}^{2} \left(\omega ^{2} -u_{0}^{2} q^{2} \right)} \delta \left(\omega -\mathbf{q}\cdot\mathbf{u}_{p} \right)\,. \qquad
\end{eqnarray}

Further calculations are not possible without specifying the explicit form of $\nu _{\mathbf{p}} $ that is the Fourier transform of interaction potential
$V\left(\left|\mathbf{R}\right|\right)$
[see  \eqref{GrindEQ__1_3_}, \eqref{GrindEQ__1_6_}]. To simplify calculations it is often accepted to replace the interaction potential
$V\left(\left|\mathbf{R}\right|\right)$
by the following effective potential (see \cite{6,7})
\begin{equation} \label{GrindEQ__3_3_}
V\left(\left|\mathbf{R}\right|\right)=\nu _{0} \delta \left(\mathbf{R}\right)\,,
\qquad
\nu _{0} =\frac{4\pi a_\mathrm{sc} }{m} \,,
\end{equation}
\noindent  where $a_\mathrm{sc} $ is the so-called $s$-wave scattering length (for details see \cite{6}). Thus, we have $\nu _{\mathbf{p}} \equiv \nu _{0} $.
Taking it into account, we have $\omega _{\mathbf{p}} \equiv \omega _{p} $ and the value $\mathbf{u}_{p} $ [see (\ref{GrindEQ__1_12_}), \eqref{GrindEQ__2_12_}] can be expressed in the form
\begin{equation}
\label{GrindEQ__3_4_}
\mathbf{u}_{p}
=\frac{\mathbf{p}}{p} u_{p} \,,
\qquad
u_{p} \equiv \frac{\partial \omega _{p} }{\partial p} =\frac{\varepsilon _{p} +\nu _{0} n_{0} }{\omega _{p} } \frac{p}{m} \,.
\end{equation}

Formulae \eqref{GrindEQ__3_3_} and \eqref{GrindEQ__3_4_} enable us to perform integration in \eqref{GrindEQ__3_2_} over the angle between vectors
$\mathbf{q}$ and $\mathbf{p}$, and then $\varepsilon _{1} \left(\mathbf{q},\omega \right)$ can be represented in the form
\begin{eqnarray}
\label{GrindEQ__3_5_}
\varepsilon _{1} \left(q,\omega \right)&=&1-\frac{\nu _{0} }{4\pi ^{2} \left(\omega ^{2} -u_{0}^{2} q^{2} \right)}\int _{0}^{\infty }\frac{\rd pp^{2} }{\omega _{p}^{2} } \frac{\partial f_{p}^{0} }{\partial \omega _{p} }
 \left(\frac{p}{mu_{p} } \omega ^{2} \omega _{p} +\varepsilon _{p} u_{0}^{2} q^{2} \right)\nonumber\\
&&\times\left(n_{0} \nu _{0} +3\varepsilon _{p} \right)
\left(1-\frac{\omega }{2qu_{p} } \ln \left|\frac{\omega +qu_{p} }{\omega -qu_{p} } \right|\right).
\end{eqnarray}

After the same angle integration, the second expression of \eqref{GrindEQ__3_2_} can be written as follows:
\begin{eqnarray}
 \label{GrindEQ__3_6_}
\varepsilon _{2} \left(q,\omega \right)=-\frac{\nu _{0} \omega }{8\pi m\left(\omega ^{2} -u_{0}^{2} q^{2} \right)} \int _{0}^{\infty }\frac{\rd pp^{2} }{\omega _{p}^{2} u_{p} }  \frac{\partial f_{p}^{0} }{\partial \omega _{p} }
\left(\frac{p}{u_{p}} \omega ^{2} \omega _{p} +\varepsilon _{p} \nu _{0} n_{0} q^{2} \right)\left(n_{0} \nu _{0} +3\varepsilon _{p} \right)\theta \left(qu_{p} -\left|\omega \right|\right),
\end{eqnarray}
\noindent  where $\theta \left(x\right)$ is Heaviside step function.

We now determine, following works \cite{12a,12b}, the sound damping rate in the case of low ($T\ll \nu _{0} n_{0} $) and high ($T\gg \nu _{0} n_{0} $ but $T<T_\mathrm{c} $) temperatures. To study the case of low temperatures it is convenient to use the variable $z\equiv \omega _{p} /T$ in the integration in \eqref{GrindEQ__3_5_} and \eqref{GrindEQ__3_6_}. The variables $p$, $\varepsilon _{p} $, $u_{p} $ in terms of the variable $z$ can be expressed using the $\omega _{p} $ explicit expression \eqref{GrindEQ__1_12_} and \eqref{GrindEQ__3_4_}. After carrying out rather cumbersome but necessary calculations, the result can be summarized as follows:
\begin{eqnarray}
 \label{GrindEQ__3_7_}
\varepsilon _{1} \left(q,\omega \right)&=&1-\frac{F}{a\left(\omega ^{2} -u_{0}^{2} q^{2} \right)} \int _{0}^{\infty }\rd z \frac{\partial f^{0} \left(z\right)}{\partial z} g\left(q,\omega ,z\right)
\left[1-\frac{\omega }{2qu_{0} v\left(z\right)} \ln \left|\frac{\omega +qu_{0} v\left(z\right)}{\omega -qu_{0} v\left(z\right)} \right|\right]\,,\nonumber
\\
\varepsilon _{2} \left(q,\omega \right)&=&-\frac{\pi F}{2a\left(\omega ^{2} -u_{0}^{2} q^{2} \right)} \int _{0}^{\infty }\rd z \frac{\partial f^{0} \left(z\right)}{\partial z} g\left(q,\omega ,z\right)
\frac{\omega }{qu_{0} v\left(z\right)} \theta \big(qu_{0} v\left(z\right)-\left|\omega \right|\big)\,,
\nonumber
\\
f^{0} \left(z\right)&=&\frac{1}{\re^{z} -1} \,,
\end{eqnarray}
\noindent  where functions $g\left(q,\omega ,z\right)$ and $v\left(z\right)$ are given by:
\begin{eqnarray}
 \label{GrindEQ__3_8_}
g\left(q,\omega ,z\right)&\equiv& \frac{\sqrt{2} \sqrt{\sqrt{1+a^{2} z^{2} } -1} \left(3\sqrt{1+a^{2} z^{2} } -2\right)}{z\left(1+a^{2} z^{2} \right)}
\left[q^{2} u_{0}^{2} \left(1+a^{2} z^{2} -\sqrt{1+a^{2} z^{2} } \right)+\omega ^{2} a^{2} z^{2} \right] ,\nonumber
\\
v\left(z\right)&\equiv& \frac{\sqrt{2} \sqrt{1+a^{2} z^{2} } \sqrt{\sqrt{1+a^{2} z^{2} } -1} }{az}\, ,
\end{eqnarray}
\noindent and the following notations were introduced [recall $u_{0} $ still given by \eqref{GrindEQ__2_10_}]:
\begin{equation}
\label{GrindEQ__3_9_}
a\equiv \frac{T}{\nu _{0} n_{0} } \,,
\qquad
 F\equiv \frac{1}{4\pi ^{2} } \nu _{0} m^{2} u_{0} =\frac{2}{\sqrt{\pi } } \sqrt{n_{0} a_\mathrm{sc}^{3} } \,.
\end{equation}

Deriving the last formula of \eqref{GrindEQ__3_9_} it is necessary to use the expressions \eqref{GrindEQ__2_10_} and \eqref{GrindEQ__3_3_}. The quantity $n_{0} a_\mathrm{sc}^{3} $ is referred to as gas parameter, see  \cite{6,7}. That is, in diluted gases the following relation should be satisfied:
\begin{equation} \label{GrindEQ__3_10_}
n_{0} a_\mathrm{sc}^{3} \ll 1.
\end{equation}

It is easy to verify that the values $\varepsilon _{1} \left(q,\omega \right)$ and $\varepsilon _{2} \left(q,\omega \right)$ can be represented as functions of one dimensionless variable $s\equiv {\omega \left(q\right)\mathord{\left/ {\vphantom {\omega \left(q\right) u_{0} q}} \right. \kern-\nulldelimiterspace} u_{0} q} $
\begin{eqnarray}
  \label{GrindEQ__3_11_}
&&\varepsilon _{1} \left(s\right)=1-\frac{F}{a\left(s^{2} -1\right)} \int _{0}^{\infty }\rd z \frac{\partial f^{0} \left(z\right)}{\partial z} g\left(s,z\right)
\left[1-\frac{s}{2v\left(z,a\right)} \ln \left|\frac{s+v\left(z,a\right)}{s-v\left(z,a\right)} \right|\right]\,,\nonumber
\\
&&\varepsilon _{2} \left(s\right)=-\frac{\pi Fs}{2a\left(s^{2} -1\right)} \int _{0}^{\infty }\rd z \frac{\partial f^{0} \left(z\right)}{\partial z} \frac{g\left(s,z\right)}{v\left(z,a\right)}
\theta \big(v\left(z,a\right)-\left|s\right|\big)\,,
\end{eqnarray}
\noindent where
\begin{equation}
 \label{GrindEQ__3_12_}
g\left(s,z\right)\equiv \frac{\sqrt{2} \sqrt{\sqrt{1+a^{2} z^{2} } -1} \left(3\sqrt{1+a^{2} z^{2} } -2\right)}{z\left(1+a^{2} z^{2} \right)}
\left(1+a^{2} z^{2} -\sqrt{1+a^{2} z^{2} } +s^{2} a^{2} z^{2} \right) ,
\end{equation}
\noindent and for these functions the symmetry conditions are valid
\begin{equation} \label{GrindEQ__3_13_}
\varepsilon _{1} \left(s\right)=\varepsilon _{1} \left(-s\right),
\qquad
\varepsilon _{2} \left(s\right)=\varepsilon _{2} \left(-s\right).
\end{equation}

Expressions \eqref{GrindEQ__2_22_} and \eqref{GrindEQ__2_23_} that determine the dispersion and decrement (or increment) of small oscillations in the system investigated, can be written in the following form by taking into account
 \eqref{GrindEQ__3_11_}:
\begin{equation}
  \label{GrindEQ__3_14_}
\varepsilon _{1} \left(s_{0} \right)=0\,,
\qquad
 s_{0} \equiv \frac{\omega _{0} \left(q\right)}{u_{0} q} \,,
\qquad
\gamma _{q} =u_{0} q\left[\varepsilon _{2} (s)\left(\frac{\partial \varepsilon _{1} (s)}{\partial s} \right)^{-1} \right]_{s=s_{0} } .
\end{equation}

As is easily seen, these oscillations have linear spectrum. In virtue of \eqref{GrindEQ__3_11_} and \eqref{GrindEQ__3_14_} the structure of dispersion equation is such that the unknown value $s$, which determines the oscillation frequency as a function of wave vector, does not depend on wave vector itself.

In accordance with \eqref{GrindEQ__2_22_} and \eqref{GrindEQ__3_14_}, the dependence of oscillation frequency $\omega _{0} $ on wave vector $q$ in this system should be determined by the solution of equation
\begin{equation}
 \label{GrindEQ__3_15_}
s_{0}^{2} -1=\frac{F}{a} \int _{0}^{\infty }\rd z \frac{\partial f^{0} \left(z\right)}{\partial z} g\left(s_{0} ,z\right)
\left[1-\frac{s_{0} }{2v\left(z\right)} \ln \left|\frac{s_{0} +v\left(z\right)}{s_{0} -v\left(z\right)} \right|\right]\,.
\end{equation}

We note that dispersion equation \eqref{GrindEQ__3_15_} is similar to the dispersion equation of zero sound in a normal Fermi liquid, (compare e.g., with the corresponding formula in  \cite{25}). Further, taking  into account this equation, to calculate the derivative ${\partial \varepsilon _{1} \left(s\right)}/{\partial s} $ appearing in \eqref{GrindEQ__3_14_}, the damping rate $\gamma _{q} $ can be presented as follows:
\begin{equation} \label{GrindEQ__3_16_}
\frac{\gamma _{q} }{u_{0} q} =-\frac{\pi }{2} \frac{Fs_{0} }{a\, b\left(s_{0} \right)} \int _{0}^{\infty }\rd z \frac{\partial f^{0} \left(z\right)}{\partial z} \frac{g\left(s_{0} ,z\right)}{v\left(z\right)} \theta \big(v\left(z\right)-\left|s_{0} \right|\big),
\end{equation}
\noindent  where
\begin{eqnarray}
  \label{GrindEQ__3_17_}
b\left(s_{0} \right)\equiv \left[\left(s^{2} -1\right)\frac{\partial \varepsilon _{1} \left(s\right)}{\partial s} \right]_{s=s_{0} }
\!\!\!\!\!&=&\!\!\!\!\frac{s_{0}^{2} +1}{s_{0} } +\frac{F}{a} \int _{0}^{\infty }\rd z \frac{\partial f^{0} \left(z\right)}{\partial z} \left[\frac{g\left(s_{0} ,z\right)}{s_{0} } -\frac{\partial g\left(s_{0} ,z\right)}{\partial s_{0} } \right]\nonumber
\\
&&\!\!\!\!+\frac{Fs_{0} }{2a} \left[\frac{\partial }{\partial s_{0} } \int _{0}^{\infty }\rd z \frac{\partial f^{0} \left(z\right)}{\partial z} \frac{g\left(s_{0} ,z\right)}{v\left(z\right)} \ln \left|\frac{s_{0} +v\left(z\right)}{s_{0} -v\left(z\right)} \right|\right].
\end{eqnarray}

We emphasize that the  above mentioned expressions \eqref{GrindEQ__3_7_}--\eqref{GrindEQ__3_17_} are exact, despite the fact that we have modified them to the form suitable to study the low temperature regime.

\begin{figure}[!t]
\begin{center}
\includegraphics[width=0.6\textwidth]{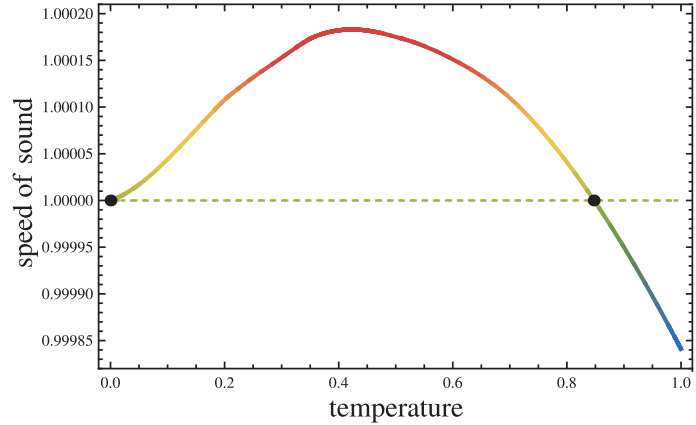}
\end{center}
\caption{(Color online) Dependence of dimensionless speed of sound $s\left(a\right) \equiv {\omega _{0} \left(q,a\right)}/{u_{0} q}$ on temperature [dimensionless quantity $a$, see \eqref{GrindEQ__3_9_}]
obtained by a numerical solution of \eqref{GrindEQ__3_15_} for $F=10^{-3} $ (solid line). Dots shows the intersection with line $s=1$ (dashed one) in points $a=0$ and $a\approx 0.847$. }
\label{fig1}
\end{figure}

As is readily seen, equation \eqref{GrindEQ__3_15_} in the general case can be solved only by numerical methods. Figure~\ref{fig1} shows the dependence $s\left(a\right)$ [see  \eqref{GrindEQ__3_11_} and \eqref{GrindEQ__3_9_}] obtained as a result of numerical solution of equation \eqref{GrindEQ__3_15_} for $F=10^{-3} $. It is evident that the function $s\left(a\right)$ behaves nonmonotonously as $a$ changes. At low $a$ region (the case of low temperature), an increase of the function is observed. At the point $a\approx 0.43$ it reaches maximum, and then $s\left(a\right)$ decreases monotonously as $a$ increases, and in the point $a\approx 0.847$ $s\left(a\right)$ it is equal to unity again. As mentioned above, the increase of the value $a$ is restricted at least to the critical temperature, see \eqref{GrindEQ__3_9_}. At zero temperature ($a=0$) we deal with a classical zero sound in Bose system, because $s=1$, and hence $\omega _{0} \left(q\right)=u_{0} q$ due to \eqref{GrindEQ__3_11_}. This result is naturally expected. However, we note once again that at the point $a\approx 0.847$ there also holds $s=1$ and the frequency of sound in dilute gas with Bose condensate is again equal to the frequency of zero sound in such a system. The fact that we discovered had not been mentioned in the literature.  This may be due to the use of the evolution equations of the system in the present work, that were obtained within the microscopic approach, as well as due to the numerical solution of the dispersion equations. Whatever the case is, the very existence of the second point of a sound dispersion curve with $s=1$, that is $a\approx 0.847$ (or $T\approx 0.847\; \nu _{0} n_{0} $) for $F=10^{-3} $, requires an individual physical interpretation. Some more comments regarding this point will be given below.

In the regions of low ($T\ll \nu _{0} n_{0} $) and high ($T\gg \nu _{0} n_{0} $ but $T<T_\mathrm{c} $) temperatures a solution of \eqref{GrindEQ__3_15_} can be expressed in analytical form. Thus, one can obtain analytical expressions for the quantity $\gamma _{q} $ in two limiting cases. To do it, as we shall see, one need to involve a numerical analysis as an auxiliary technique.

Consider first the case of low temperatures. By virtue of inequality (low temperature regime, as mentioned above)
\begin{equation} \label{GrindEQ__3_18_}
a=\frac{T}{\nu _{0} n_{0} } \ll 1
\end{equation}
\noindent  and the rapid decrease of the function $f^{0} \left(z\right)$ as $z\to \infty $  [see  \eqref{GrindEQ__3_7_}],
functions $g\left(q,\omega ,z\right)$ and $u\left(z\right)$ can be expanded in power series of $az$ in \eqref{GrindEQ__3_11_}--\eqref{GrindEQ__3_13_}:
\begin{equation}
\label{GrindEQ__3_19_}
g\left(s,z\right)\approx a^{3} z^{2} \left(s^{2} +\frac{1}{2} \right),
\qquad
 v\left(z\right)\approx 1+\frac{3}{8} a^{2} z^{2} .
\end{equation}

As will become apparent from the subsequent formulae, such an expansion corresponds to the development of perturbation theory with respect to a small parameter $a$. Taking into account the expansions \eqref{GrindEQ__3_19_}, the dispersion equation \eqref{GrindEQ__3_15_} can be written as:
\begin{eqnarray}
\label{GrindEQ__3_20_}
s_{0}^{2} -1&=&Fa^{2} \left(s_{0}^{2} +\frac{1}{2} \right)\int _{0}^{\infty }\rd z{\kern 1pt} z^{2} \frac{\partial f^{0} \left(z\right)}{\partial z}
\left[1-\frac{s_{0} }{2v\left(z\right)} \ln \left|\frac{s_{0} +1+\frac{3}{8} a^{2} z^{2} }{s_{0} -1-\frac{3}{8} a^{2} z^{2} } \right|\right] ,\nonumber
\\
 s_{0} &\equiv &\frac{\omega _{0} \left(q\right)}{u_{0} q}\, .
\end{eqnarray}

In the region of small $a$, i.e., low temperatures, as we have seen in figure~\ref{fig1}, $s\approx 1$. That is to say,  the value $\delta s\equiv s-1$ should be much less than unity, $\left|\delta s\right|\ll 1$. Therefore, the equation \eqref{GrindEQ__3_20_} admits further simplification
\begin{equation}
\label{GrindEQ__3_21_}
\delta s\approx \frac{3}{4} Fa^{2} \int _{0}^{\infty }\rd z{\kern 1pt} z^{2} \frac{\partial f^{0} \left(z\right)}{\partial z}
\left(1-\frac{1}{2} \ln 2
+\frac{1}{2} \ln \left|\delta s-\frac{3}{8} a^{2} z^{2} \right|\right) .
\end{equation}

When $F\ll 1$ and $a\ll 1$ [see  \eqref{GrindEQ__3_9_}, \eqref{GrindEQ__3_14_}] the solution of \eqref{GrindEQ__3_21_} can exist only in the $\delta s\ll a^{2} \ll 1$ domain. This inequality makes it possible to neglect the value $\delta s$ within the logarithm term in the integrand in \eqref{GrindEQ__3_21_} as the first approximation of perturbation theory. As a result, we obtain
\begin{equation} \label{GrindEQ__3_22_}
\delta s=Fa^{2} d,
\qquad
 \delta s\equiv s-1,
\end{equation}
\noindent where the notation $d$ is
\begin{equation}\begin{split}
d=\frac{3}{4} \int _{0}^{\infty }\rd z\frac{ z^{2} \re^{z} }{\left(\re^{z} -1\right)^{2} } \left(\frac{1}{2} \ln 2-1-\frac{1}{2} \ln \frac{3}{8} a^{2} z^{2} \right) .
\end{split}\end{equation}

\begin{figure}[h]
\begin{center}
\includegraphics[width=0.6\textwidth]{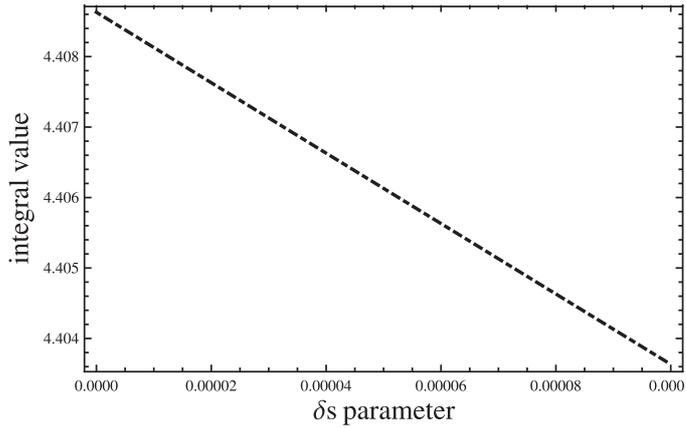}
\end{center}
\caption{Dependence $d\left(\delta s\right)$ plotted numerically according to \eqref{GrindEQ_d}  for $a=0.1$.   \label{fig2}}
\end{figure}

Formula \eqref{GrindEQ__3_22_} is also confirmed by numerical calculations. Figure~\ref{fig2} shows the dependence $d\left(\delta s\right)$, $\delta s\equiv s-1$, computed according to expression \eqref{GrindEQ__3_21_}
\begin{equation}
 \label{GrindEQ_d}
d\left(\delta s\right)
\equiv \frac{3}{4}
 \int _{0}^{\infty }\rd z\frac{ z^{2} \re^{z} }{\left(\re^{z} -1\right)^{2} }
\left(\frac{1}{2} \ln 2-1-\frac{1}{2} \ln \left|\delta s-\frac{3}{8} a^{2} z^{2} \right|\right) .
\end{equation}

\begin{figure}[h]
\begin{center}
\includegraphics[width=0.6\textwidth]{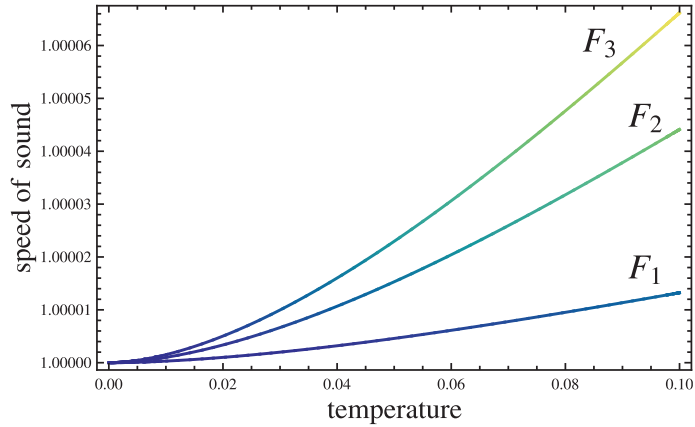}
\end{center}
\caption{(Color online) Analytical dependence of dimensionless speed of sound $s\equiv {\omega _{0} \left(q,a\right)}/{u_{0} q} $ on temperature (dimensionless quantity $a$)
in case of low temperatures,  plotted according to expressions  \eqref{GrindEQ__3_22_}, \eqref{GrindEQ__3_23_}
for $F_{1}=3 \cdot 10^{-4}$ , $F_{2}=10^{-3}$ and $F_{3}=1.5 \cdot 10^{-3}$ respectively.
\label{fig3}}
\end{figure}

As can be seen from figure~\ref{fig2}, the equality $d\left(\delta s\right)\approx d\approx 4.4$ for $a=0.1$  holds up to the second decimal. As a consequence, in the expression for $d$ that is given above, the dependence on $\delta s$ can be neglected. Then, one can derive:
\begin{equation} \label{GrindEQ__3_23_}
d\approx -\frac{\pi ^{2} }{4} \ln a-1.27,
\end{equation}
\noindent where the second term was calculated numerically, and, deriving the first term, the following integral value was used:
\[\int _{0}^{\infty }\rd z\frac{z^{2} \re^{z} }{\left(\re^{z} -1\right)^{2} }  =\frac{\pi ^{2} }{3} \,.\]
\noindent Thus, the solution \eqref{GrindEQ__3_22_} shows the following dependence of the sound frequency in dilute gas with BEC on the low temperature range:
\begin{equation} \label{GrindEQ__3_24_}
\omega _{0} \left(q\right)\approx \pm u_{0} q\left(1+Fa^{2} d\right),
\qquad
a=\frac{T}{\nu _{0} n_{0} } \ll 1.
\end{equation}

Analytical dependence $s \left(a\right) $ in case of $a \ll 1 $ in accordance with \eqref{GrindEQ__3_22_} and \eqref{GrindEQ__3_25_} is shown in figure~\ref{fig3}  for three different values of $F$.  Note that sound speed temperature corrections in a gas with BEC were also obtained in \cite{Giorgini}. In contrast to our results that are shown by formulae (\ref{GrindEQ__3_23_}), (\ref{GrindEQ__3_24_}), corrections in \cite{Giorgini} have $\delta s\sim a^{4} \ln a$ temperature dependence.

The damping rate of sound in BEC calculated according to formulae \eqref{GrindEQ__3_16_}, \eqref{GrindEQ__3_17_} and using \eqref{GrindEQ__3_19_},  \eqref{GrindEQ__3_22_}--\eqref{GrindEQ__3_24_} in the first order approximation with respect to $a$ is given by a fairly simple expression
\begin{equation} \label{GrindEQ__3_25_}
\frac{\gamma _{q} }{\omega _{0} \left(q\right)} \approx \frac{\gamma _{q} }{u_{0} q} \approx \frac{\pi ^{3} }{8} Fa^{2} =\frac{\pi ^{5/2} }{2} \sqrt{n_{0} a_\mathrm{sc}^{3} } \left(\frac{T}{\nu _{0} n_{0} } \right)^{2} .
\end{equation}

If we assume that at low temperatures [see  \eqref{GrindEQ__3_18_}] all the Bose gas is in condensate  \cite{12a,12b}, $n_{0} \left(T\right)\approx n_{0} \left(0\right)=n_{0} $,
 the quadratic dependence of decrement  $\gamma _{q} $ on temperature follows from \eqref{GrindEQ__3_24_}.
This behaviour of sound decrement at low temperatures significantly differs from that given in  \cite{1,12a,12b}.
In these papers the value $\gamma _{q} $ is known to depend on temperature in low-temperature regime  as $T^{4} $.
This circumstance,
as well as the above mentioned difference in $\delta s$ temperature dependence,
 is clearly the consequence of our using the evolution equations of the system studied,
derived in microscopic approach from the first principles.
The quadratic dependence of $\gamma _{q} $ on temperature
shows that sound damping in dilute Bose gas with condensate
can occur faster than it was expected before.

\begin{figure}[h]
\begin{center}
\includegraphics[width=0.6\textwidth]{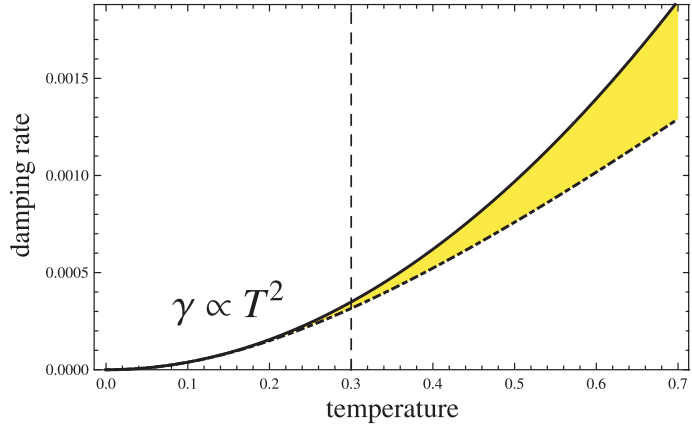}
\end{center}
\caption{(Color online) Dependence of dimensionless damping rate
$ {\gamma _{q} }/{\omega _{0}} $
  on temperature [dimensionless quantity $a$, see \eqref{GrindEQ__3_18_}]. Solid line represents the analytical approach given by \eqref{GrindEQ__3_25_}
and dotdashed line represents numerical calculation in accordance with \eqref{GrindEQ__3_15_}--\eqref{GrindEQ__3_17_}. Both lines were plotted for $F=10^{-3}$. Dashed vertical line indicates the region where analytical formula \eqref{GrindEQ__3_18_} works with good accuracy.
\label{fig4}}
\end{figure}

Figure~\ref{fig4} shows the dimensionless damping rate $\gamma _{q} / \omega _{0} \left(q\right)$ at low temperatures (low values of $a$) plotted for $F=10^{-3} $. The solid line reflects the behaviour of dimensionless decrement that follows from the analytical expression \eqref{GrindEQ__3_25_}. The dotted line shows the dependence ${\gamma _{q} \mathord{\left/ {\vphantom {\gamma _{q}  \omega _{0} \left(q\right)}} \right. \kern-\nulldelimiterspace} \omega _{0} \left(q\right)} $ on the temperature obtained as a result of numerical calculations based on formulae \eqref{GrindEQ__3_15_}--\eqref{GrindEQ__3_17_}. It is evident that analytical expression \eqref{GrindEQ__3_25_} with good accuracy coincides with the damping rate temperature dependence in a dilute gas with BEC in a rather wide range of low temperatures.

Here, we should make the following remark. The experimental data (see  \cite{3,4} and  \cite{27,28,29,30,31}) show that under the present experimental conditions, the criterion $a=\left({T\mathord{\left/ {\vphantom {T \nu _{0} n_{0} }} \right. \kern-\nulldelimiterspace} \nu _{0} n_{0} } \right)\ll 1$ is not generally  realized. For example, approximate estimates of the parameter $a$ obtained in accordance with typical experimental conditions of the mentioned studies, show the following circumstance. This parameter takes on the value $a\approx 2$ for $^{{\rm 85}} {\rm Rb}$ \cite{30}, $a\approx 23$ for $^{{\rm 23}} {\rm Na}$ \cite{4}, $a\approx 58$ for $^{{\rm 1}} {\rm H}$  \cite{27}. To perform these estimates, the scattering length values were taken from  \cite{6}. The exceptions are the data of  \cite{28} and  \cite{31}: for $^{{\rm 7}} {\rm Li}$ we have $a\approx 0.5$  \cite{28} and for $^{{\rm 133}} {\rm Cs}$ accordingly $a\approx 0.8$  \cite{31}. Therefore, to observe the effect of sound damping in dilute gas with BEC at low temperatures, the most promising are experimental conditions in  \cite{28, 31}.

Experimental conditions in other studies are rather close to the limiting case contrary to \eqref{GrindEQ__3_18_}, that is the case of high temperatures. For this reason, we now consider, as in  \cite{12a,12b}, another limiting case
\begin{equation} \label{GrindEQ__3_26_}
a\equiv \frac{T}{\nu _{0} n_{0} } \gg 1,
\qquad
 T<T_\mathrm{c} \,.
\end{equation}

As before, we assume here that $\varepsilon _{p} \sim \nu _{0} n_{0} $. Then, from \eqref{GrindEQ__3_26_} it follows that $\omega _{p} \left(n_{0} \right)\ll T$.
In this case, the following limiting expression for the distribution function $f_{p}^{0} \left(\omega _{p} \right)$ can be used (see, in this regard \eqref{GrindEQ__2_2_} and \cite{12a,12b})
\begin{equation} \label{GrindEQ__3_27_}
f_{p}^{0} \left(\omega _{p} \right)\approx \frac{T}{\omega _{p} } ,
\qquad
 \varepsilon _{p} \sim \nu _{0} n_{0} \,,
\qquad
 \omega _{p} \left(n_{0} \right)\ll T\,.
\end{equation}

To simplify the further calculations in \eqref{GrindEQ__3_5_} and \eqref{GrindEQ__3_6_}, it is convenient to change the integration variable $p$ to variable $z$, that is introduced in accordance with the formula
\[\frac{p^{2} }{2m\nu _{0} n_{0} } =z\,.\]

As a result of the change of integration variable, we obtain the following expressions:
\begin{equation}\label{GrindEQ__3_28_}
\varepsilon _{1} \left(s\right)=1-\frac{Fa}{\left(s^{2} -1\right)} \int _{0}^{\infty }\rd z \, g\left(s,z\right)
\left[1-\frac{s}{2v\left(z\right)} \ln \left|\frac{s+v\left(z\right)}{s-v\left(z\right)} \right|\right],
\end{equation}
\noindent  where now we have introduced the notations
\begin{equation} \label{GrindEQ__3_29_}
g\left(s,z\right)\equiv \frac{\sqrt{2} \left(3z+1\right)\left[z\left(s^{2} +1\right)+2s^{2} +1\right]}{\sqrt{z} \left(z+1\right)\left(z+2\right)^{2} } \,,
\qquad
v\left(z\right)\equiv \sqrt{2} \frac{z+1}{\sqrt{z+2} }\, ,
\end{equation}
and the quantities $F$ and $a$ are still defined by \eqref{GrindEQ__3_9_} and \eqref{GrindEQ__3_10_}.

From the condition $\varepsilon _{1} \left(s_{0} \right)=0$ [see (2.19), \eqref{GrindEQ__2_23_}, \eqref{GrindEQ__3_24_}, \eqref{GrindEQ__3_28_}], written when $s_{0} \left(a\right)\approx \pm \left[1+\delta s\left(a\right)\right]$, $\left|\delta s\right|\ll 1$, in the form of
\begin{eqnarray} \label{GrindEQ__3_30_}
&&\delta s=Fa\int _{0}^{\infty }\rd z\frac{\left(3z+1\right)\left(2z+3\right)}{\sqrt{2z} \left(z+1\right)\left(z+2\right)^{2} }
\left[1-\frac{1}{2v\left(z\right)} \ln \left|\frac{1+v\left(z\right)+\delta s}{1-v\left(z\right)+\delta s} \right|\right] ,\nonumber\\
&&\delta s\left(a\right)\equiv s\left(a\right)-1\,,
\end{eqnarray}
\noindent it follows that equation \eqref{GrindEQ__3_24_} has a solution if only $Fa\ll 1$ [but $a\gg 1$ see \eqref{GrindEQ__3_26_}]. There is no analytical solution to this equation. However, the numerical analysis reveals that the result of integration with respect to $z$ in the right-hand side of formula \eqref{GrindEQ__3_30_} almost does not depend on $\delta s$ if $\left|\delta s\right|\ll 1$. In this regard, the solution of \eqref{GrindEQ__3_30_} in case of $\left|\delta s\right|\ll 1$ can be written as follows:
\begin{equation} \label{GrindEQ__3_31_}
\delta s=-Fad_{2} \,,
\qquad
 Fa\ll 1\,,
\end{equation}
\noindent where the constant $d_{2} $ is given by
\begin{equation}
\label{GrindEQ__3_31_a}
d_{2} \equiv \int _{0}^{\infty }\rd z\frac{\left(3z+1\right)\left(2z+3\right)}{\sqrt{2z} \left(z+1\right)\left(z+2\right)^{2} }
\left[1-\frac{\sqrt{z+2} }{2\sqrt{2} \left(z+1\right)} \ln \left|\frac{\sqrt{z+2} +\sqrt{2} \left(z+1\right)}{\sqrt{z+2} -\sqrt{2} \left(z+1\right)} \right|\right] \approx 3.97.
\end{equation}

\begin{figure}[h]
\begin{center}
\includegraphics[width=0.6\textwidth]{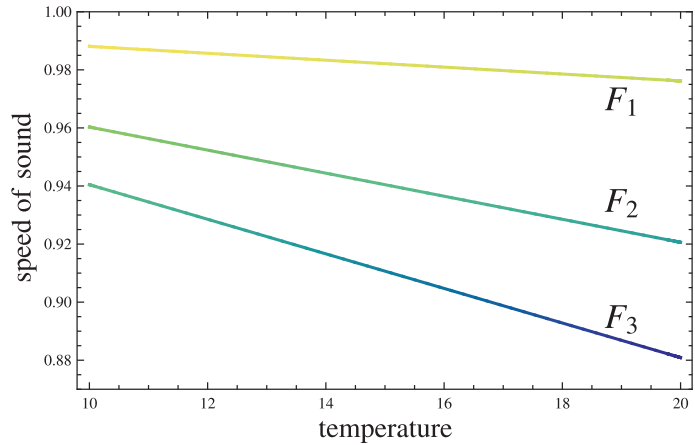}
\end{center}
\caption{(Color online)  Analytical dependence of the dimensionless sound speed  $s\equiv {\omega _{0} \left(q,a\right)}/{u_{0} q} $ on temperature (dimensionless quantity $a$)
in case of high temperatures,  plotted according to expressions  \eqref{GrindEQ__3_31_}, \eqref{GrindEQ__3_31_a}
for $F_{1}=3 \cdot 10^{-4}$ , $F_{2}=10^{-3}$ and $F_{3}=1.5 \cdot 10^{-3}$ respectively.
\label{fig5}}
\end{figure}

Analytical dependence $s \left(a\right) $ in case of $a \gg 1 $ in accordance with \eqref{GrindEQ__3_31_} and \eqref{GrindEQ__3_31_a} is shown in figure~\ref{fig5}  for three different values of $F$.

The value $b\left(s\right)$ required in accordance with \eqref{GrindEQ__3_16_}, \eqref{GrindEQ__3_17_} to calculate the damping rate of sound at high temperatures [see \eqref{GrindEQ__3_26_}] is given by:
\begin{eqnarray} \label{GrindEQ__3_32_}
b\left(s\right)&\equiv& \left[\left(s^{2} -1\right)\frac{\partial \varepsilon _{1} \left(s\right)}{\partial s} \right]_{\omega =\omega _{0} } =
\frac{s_{0}^{2} \left(a\right)+1}{s_{0} \left(a\right)}
-\frac{Fa}{s_{0} \left(a\right)} \int _{0}^{\infty }\rd z {\kern 1pt} g\big(s_{0} \left(a\right),z\big)\frac{v^{2} \left(z\right)}{s_{0}^{2} \left(a\right)-v^{2} \left(z,a\right)}\nonumber
\\
&&-\frac{Fa}{s_{0} \left(a\right)} \int _{0}^{\infty }\rd z \frac{\partial g\big(s_{0} \left(a\right),z\big)}{\partial s_{0} \left(a\right)}
\left[1-\frac{s_{0} \left(a\right)}{2v\left(z,a\right)} \ln \left|\frac{s_{0} \left(a\right)+v\left(z,a\right)}{s_{0} \left(a\right)-v\left(z,a\right)} \right|\right],
\end{eqnarray}
\noindent where the functions $g\left(s,z\right)$and $v\left(z\right)$ are still defined by \eqref{GrindEQ__3_29_}, and $s_{0} \left(a\right)$ is given by \eqref{GrindEQ__3_30_} while taking  \eqref{GrindEQ__3_31_} into account. One can verify that the first term only yields the main contribution to $b\left(s\right)$.
The damping rate $\gamma _{q} $ of sound in BEC, calculated according to formulae \eqref{GrindEQ__3_16_} using \eqref{GrindEQ__3_32_}, in the main approximation with respect to $Fa$ [see  \eqref{GrindEQ__3_31_}] indicates a linear dependence on temperature:
\begin{equation} \label{GrindEQ__3_33_}
\frac{\gamma _{q} }{\omega _{0} \left(q\right)} \approx \frac{\gamma _{q} }{u_{0} q} \approx \frac{\pi \left(\pi +6\right)}{8} Fa,
\end{equation}
and that is a well known result (see for example \cite{2,12a,12b}). Deriving \eqref{GrindEQ__3_33_}, the following integral was used
\[\int _{0}^{\infty }\rd z\frac{\left(3z+1\right)\left(2z+3\right)}{\sqrt{z} \left(z+1\right)^{2} \left(z+2\right)^{3/2} }  =\frac{\pi +6}{2} \,.\]
Formula \eqref{GrindEQ__3_33_} using \eqref{GrindEQ__2_10_},  \eqref{GrindEQ__3_3_},  \eqref{GrindEQ__3_9_} can be written as
\begin{equation} \label{GrindEQ__3_34_}
\gamma _{q} =\frac{\pi +6}{8} a_\mathrm{sc} Tq\,.
\end{equation}

It should be mentioned that this expression differs prima facie from that given in \cite{2,12a,12b}, where
\begin{equation} \label{GrindEQ__3_35_}
\gamma _{q} =\frac{3\pi }{8} a_\mathrm{sc} Tq\,.
\end{equation}

However, it is easy to verify that the numerical factors in \eqref{GrindEQ__3_34_} and \eqref{GrindEQ__3_35_} coincide with two significant digits. Note that in \cite{12a,12b} it was also indicated that the formula (4.38) obtained therein  slightly differs in a numerical factor from a similar formula of the previous studies. In our case, the method used to obtain formula (\ref{GrindEQ__3_34_}) differs greatly from the approaches used in \cite{12a,12b} to derive (\ref{GrindEQ__3_35_}). This is what really caused the difference (although small) in formulae (\ref{GrindEQ__3_34_}) and (\ref{GrindEQ__3_35_}).

\begin{figure}[h]
\begin{center}
\includegraphics[width=0.6\textwidth]{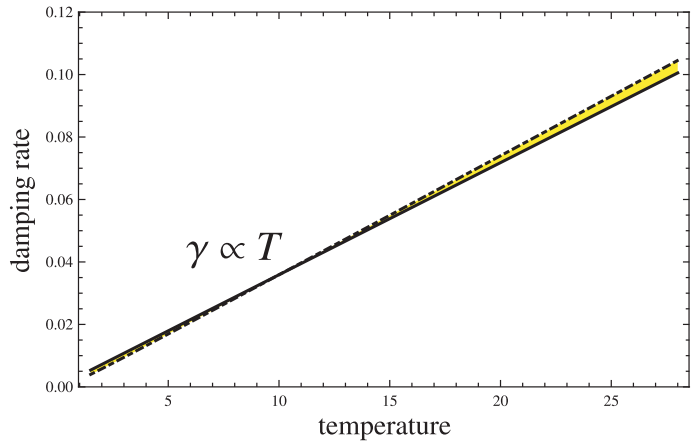}
\end{center}
\caption{(Color online) Dependence of dimensionless damping rate
${\gamma _{q} }/{\omega _{0}} $
on temperature [dimensionless quantity $a$, see \eqref{GrindEQ__3_26_}] in the case of high temperatures.
The solid line shows the dependence given by  \eqref{GrindEQ__3_33_}  and the dashed one reveals the result of numerical computing based on the formulae \eqref{GrindEQ__3_15_}--\eqref{GrindEQ__3_17_}.
Both lines were plotted for $F=10^{-3}$.
\label{fig6}}
\end{figure}

Figure~\ref{fig6} shows the dependence of dimensionless damping rate ${\gamma _{q}/ \omega _{0} \left(q\right)} $ at high temperature region [$a\equiv \left({T/\nu _{0} n_{0} } \right)\gg 1$ and $T<T_\mathrm{c} $] plotted for $F=10^{-3} $. It is apparent that the graph plotted according to formula \eqref{GrindEQ__3_33_} or \eqref{GrindEQ__3_34_} (solid line) nearly coincides with the graph obtained by numerical computations based on formulae \eqref{GrindEQ__3_15_}--\eqref{GrindEQ__3_17_} (dashed line). As it follows from figure~\ref{fig4}, the condition of sound existence in a dilute gas with BEC, $\left({\gamma _{q} /\omega _{0} } \right)\ll 1$, is well satisfied even in the region where $a\sim 30$. Naturally, one should be confident that in a particular physical system, the condition $T<T_\mathrm{c} $ is also satisfied in this range of $a$.

\section{Conclusion}

We have reported the main results related to the construction of microscopic theory of  sound damping due to Landau mechanism in a dilute gas with Bose condensate.
The analytical expressions of propagation velocity and damping rate of sound in a dilute gas with Bose condensate in the limiting cases of high and low temperatures were obtained. It was shown that at high temperatures these expressions coincide with those obtained previously by other authors in various phenomenological approaches. At low temperatures, the behaviour of collisionless sound decrement obtained in the present paper, significantly differs from the same decrement obtained by other authors. In our opinion, the distinction is caused by our use of evolution equations that were obtained in the microscopic approach from the first principles.

We now make a significant remark. At first glance, the presence of general equations \eqref{GrindEQ__3_14_}--\eqref{GrindEQ__3_17_} makes it possible to find the parameters of propagation and damping of sound in the present system in any temperature range (i.e., for all values $a$), at least numerically. Meanwhile, figure~\ref{fig3} and figure~\ref{fig4} do not display the data for the `intermediate' temperature range, that is from $a=0.7$ to $a=1.4$. This is caused by the following circumstance. We have already mentioned that in this range of values of $a$ there are no analytic methods for solving dispersion equations \eqref{GrindEQ__3_14_}--\eqref{GrindEQ__3_17_}. But as it turned out, in this temperature range, the numerical methods, at least those we have used, also become uncontrollable. Hence, one cannot trust its results. This is supposed to be due to the fact that in this interval of $a$, the point $a\approx 0.847$ is located. Recall that it is the point where the value of $s_{0} \left(a\right)={\omega _{0} \left(q,a\right)\mathord{\left/ {\vphantom {\omega _{0} \left(q,a\right) u_{0} q}} \right. \kern-\nulldelimiterspace} u_{0} q} $ [see \eqref{GrindEQ__3_14_}] is equal to unity. As it was mentioned earlier, the same value  $s$ gains also at the point $a=0$ (zero sound). It is that very neighbourhood of $a\approx 0.847$ that gives  huge oscillations of sound damping rate in the received outcome of numerical computations with insignificant changes of the value $a$. However, one cannot be sure that these oscillations of the damping rate reflect a real physical picture. The reason for doubting the outcome lies in the mentioned uncertainty in this temperature range of the numerical methods used in the present paper. Numerical calculations are poorly controlled due to the slow convergence of integrals in \eqref{GrindEQ__3_15_}--\eqref{GrindEQ__3_17_} as the manifestation of non-analytic dependence on temperature of the dispersion characteristics of the system in the neighbourhood of the point $a\approx 0.847$. Apparently, a more detailed study of the behaviour of a decrement (or maybe even an increment!) of sound in the neighbourhood of this point requires the use of more sophisticated numerical methods. The authors intend to address this issue soon.

\begin{figure}[!b]
\begin{center}
\includegraphics[width=0.6\textwidth]{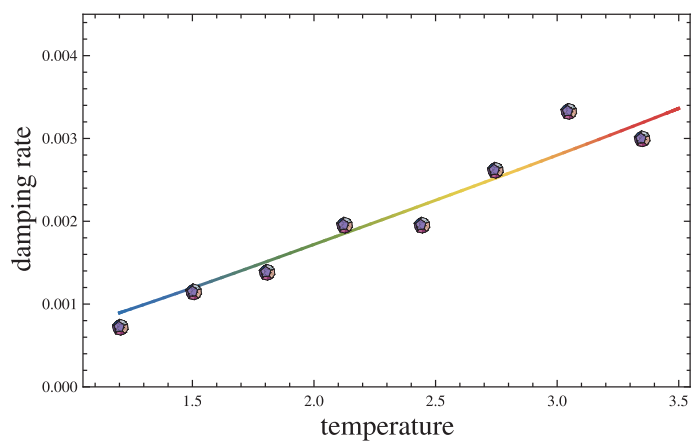}
\end{center}
\caption{(Color online) The dependence of dimensionless damping rate
$ {\gamma _{q} }/{\omega _{0}} $
on temperature (dimensionless quantity $a$). The solid line displays the result of numerical computation for $F=3\cdot 10^{-4}$ based on expressions  \eqref{GrindEQ__3_15_}--\eqref{GrindEQ__3_17_}. Diamonds reproduce the experimental points taken from~\cite{28}.
\label{fig7}}
\end{figure}

In this regard, the results in \cite{32} are also of interest, where propagation and absorption of the transverse breathing mode of an elongated BEC were investigated experimentally in $^{87}$Rb vapour. The authors of the mentioned study, attempting to measure the damping rate of these modes at the temperature approximately equal to $40\div60$~nK, had found that the behaviour of the perturbation amplitude in this temperature region differs significantly from the behaviour of the amplitude of a damped sinusoidal signal. The presence of such a phenomenon was suggested to be explained due to nonlinear effects in the propagation of the modes studied in the system. We also would like to draw attention to the fact that temperature range $40\div60$~nK corresponds to the values of quantity $a$ in terms of the present paper that are in the $0.7\div1.0$ range. The value of $a$ is naturally calculated according to formula \eqref{GrindEQ__3_9_} and using the values of the physical characteristics of the system \cite{32}. In other words, in the mentioned case \cite{32}, it deals with the nearest neighbourhood of the point $a\approx 0.847$ where the nonanalytic dependence of the dispersion characteristics on temperature is revealed.

In conclusion, we note a good correlation of our theory in the case of high temperature region with the experimental data \cite{32}. As can be seen from figure~\ref{fig7}, the results of numerical calculations based on formulae \eqref{GrindEQ__3_15_}--\eqref{GrindEQ__3_17_} for $F=3\cdot 10^{-4} $ [see \eqref{GrindEQ__3_9_}] are in satisfactory concordance with the experimental data \cite{32}. Experimental data reveal the same good fit with the dependence given by formula \eqref{GrindEQ__3_33_}.

The authors are naturally aware of the fact that a direct comparison of the results of the present paper with the experimental data \cite{32} can hardly be considered entirely correct. At least, the reason is that the authors of \cite{32} have to deal with the trapped BEC. However, such a comparison once again demonstrates the validity of the statement \cite{12a,12b} that the linear dependence of the sound damping rate on temperature should occur in the case of trapped BEC.

Noting the qualitative similarity in particular cases of our results with the experimental ones, we especially emphasize once again
that we do not claim to  microscopically describe the damping of sound in trapped BEC.
This specific case requires a cardinal modification of the whole theory,
and it is the issue the authors are currently working on.


\ukrainianpart

\title{Про беззіткненнєвий механізм загасання звуку в розріджених бозе-газах із конденсатом}
\author{Ю.В. Слюсаренко\refaddr{label1,label2}, О.Ю. Крючков\refaddr{label2}}
\addresses{
\addr{label1}Інститут теоретичної фізики ім.~О.І.~Ахієзера, ННЦ ХФТІ, вул. Академічна, 1, 61108 Харків, Україна
\addr{label2}Харківський національний університет ім.~В.Н.~Каразіна, пл. Свободи, 4,  61077 Харків, Україна
}
%
%
%

\makeukrtitle

\begin{abstract}
\tolerance=3000%
Побудовано мікроскопічну теорію загасання звуку за механізмом Ландау у розріджених газах із бозе-конденсатом. В основу теорії було закладено пов'язані рівняння еволюції для парамерів опису системи. Ці рівняння було виведено у більш ранніх роботах у мікроскопічному підході, що базується на викорис\-танні методу скороченого опису квантових систем багатьох частинок (метод Пелетминського) та моделі Боголюбова для слабко неідеального бозе-газу з виділеним конденсатом. Отримано рівняння дисперсії звукових коливань у системі, що вивчається, шляхом лінеаризації зазначених рівнянь еволюції у без\-зіткненнєвому  наближенні. Проведено аналіз рівнянь дисперсії, як чисельно, так і аналітично. Одержано аналітичні вирази для швидкості розповсюдження й коефіцієнта поглинання звуку в розріджених газах із бозе-конденсатом у граничних випадках великих та малих температур. Нами продемонстровано, що в області малих температур температурна залежність знайдених величин суттєво відрізняється від тих, що отримані раніше іншими авторами в напівфеноменологічних підходах. Вказано на можливі ефекти, пов'язані з неаналітичними залежностями дисперсійних характеристик системи від температури.
\keywords розріджений бозе-газ, бозе-ейнштейнівська конденсація (БЕК),
мікроскопічна теорія, звук, механізм Ландау, дисперсійні співвідношення,
швидкість звуку, показник загасання

\end{abstract}

\end{document}